# A Showdown of ChatGPT vs DeepSeek in Solving Programming Tasks

Ronas Shakya
*Centre for the Science of Learning & Technology (SLATE)*
*Univeristy of Bergen*
Bergen, Norway
ronas.shakya@uib.no

Farhad Vadiee
*Centre for the Science of Learning & Technology (SLATE)*
*Univeristy of Bergen*
Bergen, Norway
farhad.vadiee@uib.no

Mohammad Khalil
*Centre for the Science of Learning & Technology (SLATE)*
*Univeristy of Bergen*
Bergen, Norway
mohammad.khalil@uib.no

*Abstract*—The advancement of large language models (LLMs) has created a competitive landscape for AI-assisted programming tools. This study evaluates two leading models: ChatGPT 03-mini and DeepSeek-R1 on their ability to solve competitive programming tasks from Codeforces. Using 29 programming tasks of three levels of easy, medium, and hard difficulty, we assessed the outcome of both models by their accepted solutions, memory efficiency, and runtime performance. Our results indicate that while both models perform similarly on easy tasks, ChatGPT outperforms DeepSeek-R1 on medium-difficulty tasks, achieving a 54.5% success rate compared to DeepSeek's 18.1%. Both models struggled with hard tasks, thus highlighting some ongoing challenges LLMs face in handling highly complex programming problems. These findings highlight key differences in both model capabilities and their computational power, offering valuable insights for developers and researchers working to advance AI-driven programming tools.

*Keywords—ChatGPT, DeepSeek, AI-assisted programming, programming education, Large language models (LLMs)*

## I. INTRODUCTION

The rapid emergence of artificial intelligence (AI) technologies has sparked a surge in research focused on evaluating code generation capabilities. Large Language Models (LLMs) are becoming increasingly reliable for a wide range of users, including novice programmers and experienced developers. This accessibility presents a significant opportunity to scale up computing education, potentially transforming both self-paced learning environments and formal higher education settings. LLMs offer promising applications in automating assignment generation and provide personalized feedback as well as supporting conceptual understanding in programming courses. The well-known model ChatGPT, known for its ability to generate human-like text, has firmly established itself as a household name. ChatGPT has rapidly integrated into different parts of our society, providing support in education for students and teachers, including a significant contribution in programming. The iterations of ChatGPT continue to progress with newer models being developed such as ChatGPT o3-mini and ChatGPT-4o. These new models offer enhanced accuracy, faster response times, and improved ability to handle complex queries [15]. In particular, ChatGPT o3-mini has left a positive impact in the community via its high-performance results on science and coding tasks [16]. Recently, ChatGPT o3-mini has become widely favored by developers for its higher rate limits and lower latency, making it a compelling choice for coding and logical problem-solving tasks. Similarly, the ChatGPT-4o presents strong performance in multimodal capabilities and various context handling.

However, the AI landscape has witnessed intense competition among global tech giants in the area of AI technologies, particularly for LLMs. A notable recent challenger has emerged from China called DeepSeek, an open source chatbot designed with a primary focus on logical reasoning and deductive thinking, to solve complex tasks. On January 20, 2025, DeepSeek released its latest models, DeepSeek-R1, which has garnered significant attention in headlines for achieving comparable and remarkable results like ChatGPT while requiring only a fraction of the training cost [6]. DeepSeek-R1's impressive results include scoring 79.8% on the AIME 2024 math benchmark, slightly outperforming ChatGPT o1's 79.2% [14].

Building on the remarkable performance of DeepSeek-R1 in contrast to ChatGPT models, this study seeks to evaluate and compare the capabilities of DeepSeek-R1 and ChatGPT o3-mini in the context of solving coding tasks. Specifically, we investigate how these models perform in terms of accuracy, efficiency, and execution quality. By quantifying both models' performance, we aim to assess whether emerging LLM models like DeepSeek can challenge ChatGPT's dominance in AI-assisted coding workflows, therefore, offering valuable insights into their potential impact on developers' decision-making processes and the future of AI-enhanced software development.

## II. BACKGROUND

The integration of AI in programming has transformed the learning scenario in programming, thus offering benefits to students, educators and professional developers. In particular, LLMs have emerged as powerful AI tools in the field of programming by leveraging deep neural networks trained on vast amounts of textual data. When these models are trained on extensive codebases, they further gain an ability to generate human-like code from natural language prompts. This capability has led to the development of specialized LLMs for programming tasks, such as ChatGPT o3-mini, GitHub Copilot, and Google PaLM, which are fine-tuned versions of more general models like ChatGPT3. These advancements have led to an increase in research aimed at improving LLMs for programming, focusing on performance optimization, cost efficiency, and accessibility [24,21].

The use of LLMs in programming has created new opportunities for both novice and experienced developers.

LLM models can streamline repetitive coding tasks, accelerate prototyping, and clarify complex algorithms [11]. For beginners, LLMs may serve as interactive learning tools by offering explanations and generating code samples that enhance their understanding of programming concepts [13]. Experienced developers can leverage LLMs to optimize code efficiency, explore alternative implementations, and quickly adapt to new programming languages or frameworks [23].

There are recent research works that have focused on evaluating the reliability of LLM-generated code. Studies have examined various aspects, including the accuracy of code generation, the model's ability to understand programming concepts, and its performance across different programming languages. For example, Ramler et al. explored the application of ChatGPT in undergraduate computer science curricula, offering opportunities and challenges in teaching fundamental programming courses [20]. Similarly, Chen et al. introduced StuGPTViz, a visual analytics system that tracks and compares temporal patterns in student prompts and the quality of ChatGPT's responses which offered significant pedagogical insights for instructors [1]. Another example is Copilot, powered by GPT-4, which integrates with popular code editors to provide AI-driven coding assistance. One Study revealed that programmers using Copilot completed tasks approximately 55.8% faster than those without AI assistance, highlighting the potential impact of LLMs on software development productivity [16]. Other research from the literature investigated ChatGPT's capabilities in Python code generation suggests that they can effectively aid novice programmers in solving complex coding challenges using minimal prompts [2].

DeepSeek, on the other hand, has demonstrated remarkable reasoning abilities in challenging problem-solving tasks. For instance, [4] reported that DeepSeek-R1 outperformed ChatGPT and Gemini on a set of 30 challenging mathematical problems from the MATH dataset. Another study by [5] reported that DeepSeek outperformed three widely used large language models: Gemini, GPT, and Llama, but lagged behind Claude in machine learning classification tasks. [5] also noted that while DeepSeek operates at a slower speed compared to these models, its open-source nature and cost efficiency offer significant advantages to users.

However, studies also indicate that human intervention is often necessary to guide AI-generated solutions in the right direction [2, 18]. While this study does not primarily focus on evaluating LLMs without human intervention, our goal is to assess the potential of the new Chinese model, DeepSeek-R1, alongside OpenAI's ChatGPT in solving programming tasks with a focus on accuracy of results, memory usage, and time consumption to provide solutions. According to recent research, LLMs are becoming more and more important in software development, with applications in debugging, code completion, and algorithm optimization [24]. Despite the promising capabilities of LLMs, especially those pretrained on code, there is still a notable lack of comprehensive evaluations of their language-to-code generation performance [25]. By addressing this gap, our study provides valuable insights into how well these models perform in actual coding situations and advances our knowledge of AI-assisted programming.

## III. METHODOLOGY

### A. Coding tasks

In our study, we have selected 29 tasks from Codeforces (https://codeforces.com/). Codeforces is one of the largest and most widely used competitive programming platforms (i.e., sport programming competitions) which provides an open infrastructure for organizing and conducting programming contests, as well as automating training courses in algorithmic problem-solving [11]. Codeforces offer a large set of programming languages contests, including C++, Python, and Java (see Fig. 1).

The selected tasks in Codeforces are categorized based on their difficulty levels into three groups: easy, medium, and hard programming tasks. The difficulty level of a programming problem is determined by its complexity and the frequency of successful attempts by users. For instance, a problem that has been solved by a large number of individuals with relatively few attempts is classified as easy, while tasks requiring more attempts or exhibiting lower success rates are categorized as medium or hard accordingly. For our study, we used C++ as the programming language to prompt the two LLMs due to its fast compilation speed and widespread use in competitive programming.

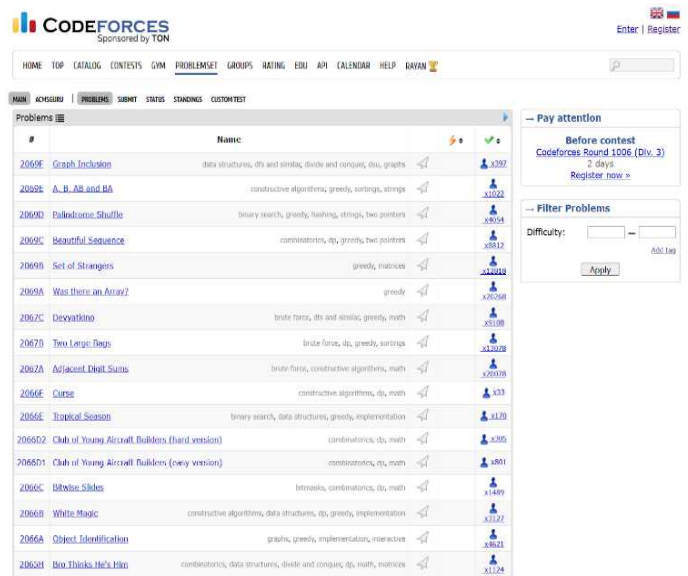

Fig. 1. Codeforces for competitive programming

### B. Study setting

ChatGPT-o3-mini was used to generate the code for ChatGPT, while DeepSeek-R1 model was used for DeepSeek. On Codeforces, multiple compiler options are available for C++ submissions, including versions of GNU G++, Clang, and MSVC. To ensure a fair comparison and uniform execution across all solutions, we selected GNU G++20 13.2 (64-bit, winlib) as the compiler. This choice helps standardize performance evaluation by minimizing variations in compilation behavior, optimizations, and runtime performance that different compilers might introduce. Each programming problem was presented to the LLM models using the following consistent prompt:

*"Solve the following question and return C++ code, which is compilable by g++ 20. Pay attention to memory and time limits."* followed by the selected tasks from the Codeforces with input and output examples (see Fig .2.). For each model we collected 29 solutions, which were compiled and compared.

After the prompts are used in the models, the codes are generated by the LLMs, we submit them to Codeforces, which then provides feedback such as *'Accept', 'Wrong Answer', 'Memory Limit Exceeded', 'Time Limit Exceeded',* and *'Compilation Error'*. If the feedback is *'Accept'*, the code is considered correct otherwise, it is categorized as wrong.

Fig. 2. Example prompt for generating coding solutions to one of the easy programming tasks.

### C. Data Analysis

We calculated the percentage of accepted programming tasks for each model. To assess overall performance, we used a weighted scoring formula that considers accepted tasks, execution efficiency, memory usage, and time consumption. Our weighted scoring formula that we use to evaluate the models:

$$Weighted\ score = X \times (Difficulty\ score \times (1 - \alpha \times TimeNormalized - \beta \times MemoryNormalized))\ \ldots\ldots (I)$$

Where,
X = Accepted solutions
Difficulty score = 1 (Easy), 2 (Medium), and 3 (Hard),
α (alpha) = penalty factor for execution time = 0.5,
β (beta) = penalty factor for memory usage = 0.5

In the equation (I), we set the penalty factors α (alpha) and β (beta) to 0.5 to give equal importance to execution time and memory usage. This will keep the balance and maintain the fair evaluation of the solutions.

## IV. RESULTS

Our analysis revealed interesting findings regarding solution acceptance, execution efficiency, and memory consumption for the programming tasks. Table 1 presents a comparative analysis of the results of the ChatGPT and DeepSeek LLMs in solving the 29 programming tasks of varying difficulty levels.

TABLE I. RESULTS ON DIFFERENT DIFFICULTY LEVELS OF THE SELECTED PROGRAMMING TASKS WITH THEIR MEMORY USAGE AND TIME CONSUMPTION.

| Question number | Difficulty level | ChatGPT (o3-mini) | Memory | Time | DeepSeek (R1) | Memory | Time |
|---|---|---|---|---|---|---|---|
| 1 | Easy | Accept | 0 | 62 | Accept | 52 | 77 |
| 2 | Easy | Accept | 100 | 124 | Accept | 100 | 154 |
| 3 | Easy | Accept | 100 | 124 | Accept | 100 | 124 |
| 4 | Easy | Accept | 0 | 92 | Accept | 100 | 62 |
| 5 | Easy | Accept | 0 | 124 | Accept | 100 | 124 |
| 6 | Easy | Accept | 100 | 154 | Accept | 0 | 124 |
| 7 | Easy | Accept | 100 | 77 | Wrong answer | 0 | 30 |
| 8 | Easy | Accept | 0 | 156 | Accept | 100 | 186 |
| 9 | Easy | Accept | 100 | 124 | Accept | 100 | 154 |
| 10 | Medium | Time limit exceeded | 4300 | 0 | Wrong answer | 0 | 15 |
| 11 | Medium | Accept | 100 | 468 | Wrong answer | 46 | 100 |
| 12 | Medium | input idle error | 0 | 0 | Wrong answer | 0 | 15 |
| 13 | Medium | Accept | 3500 | 140 | Wrong answer | 0 | 46 |
| 14 | Medium | Accept | 100 | 77 | Wrong answer | 0 | 15 |

| 15 | Medium | Accept | 15200 | 233 | Wrong answer | 4700 | 31 |
| 16 | Medium | Accept | 100 | 124 | Accept | 200 | 124 |
| 17 | Medium | Time limit exceeded | 100 | 1000 | Accept | 3100 | 966 |
| 18 | Medium | Accept | 500 | 218 | Wrong answer | 0 | 60 |
| 19 | Medium | Wrong answer | 0 | 92 | Wrong answer | 0 | 60 |
| 20 | Medium | Wrong answer | 0 | 62 | Compilation error | 0 | 0 |
| 21 | Hard | couldn't generate | 0 | 0 | Time limit exceeded | 3600 | 4000 |
| 22 | Hard | Accept | 100 | 124 | Wrong answer | 0 | 46 |
| 23 | Hard | Wrong answer | 0 | 92 | Wrong answer | 100 | 60 |
| 24 | Hard | Wrong answer | 0 | 46 | Memory limit exceeded | 524300 | 468 |
| 25 | Hard | Wrong answer | 0 | 61 | Wrong answer | 0 | 31 |
| 26 | Hard | Wrong answer | 0 | 46 | Wrong answer | 0 | 46 |
| 27 | Hard | Runtime error | 114100 | 1577 | Compilation error | 0 | 0 |
| 28 | Hard | Wrong answer | 0 | 46 | Wrong answer | 0 | 30 |
| 29 | Hard | Wrong answer | 0 | 30 | Wrong answer | 100 | 46 |

### A. Programming Problem solving abilitythors

#### a) Performance of ChatGPT

All of the 9 easy tasks from the dataset were accepted for the code provided by the ChatGPT. For medium level tasks, only 6 out of 11 tasks were accepted, which accounts for 54.5% of acceptance rate. In the case of hard level tasks, ChatGPT managed to get only 1 out of 9 tasks accepted with most failure due to *'Wrong answers'* for most of the tasks followed by *'Runtime error'* and *'couldn't generate'*. As shown in Table 2, the accepted solutions for medium level tasks consumed more memory compared to the easy and hard level tasks. Despite this higher resource usage, Fig. 3 shows that ChatGPT achieved a significantly higher weighted score (11.6) on medium level tasks.

Overall, 44.8% of the solutions from ChatGPT were not accepted. The most common reasons for failure were *'Wrong answer'* and *'Time limit exceeded'.* This suggests the compilation took too long to execute within the given constraints.

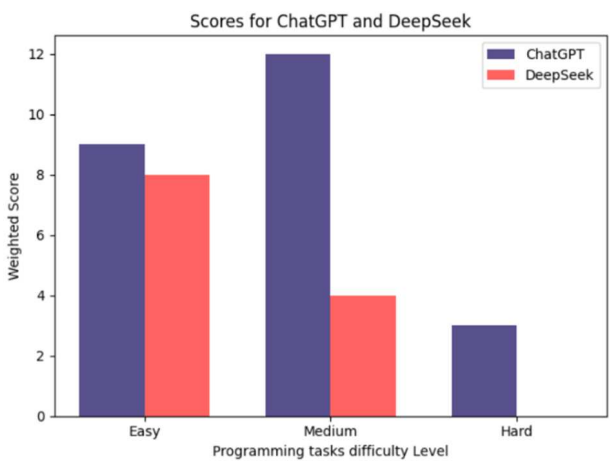

Fig.3. The bar graph shows the average weighted score for different difficulty levels of programming tasks.

#### b) Performance of DeepSeek

For easy level tasks, the performance of DeepSeek was almost similar to that of ChatGPT but for medium level tasks the success rate for DeepSeek is only 18.1% as shown in table 2. DeepSeek struggled with hard tasks, often generating incorrect or inefficient solutions with feedback such as *'Wrong answer'*, *'Memory limit exceeded'*, *'Compilation error'* and *'Time limit exceeded'.* Around 65.51% of the solutions from DeepSeek were not accepted. DeepSeek suffers from compilation errors and memory issues.

TABLE II. RESULTS ON DIFFERENT DIFFICULTY LEVELS OF THE SELECTED PROGRAMMING TASKS WITH THEIR MEMORY USAGE AND TIME CONSUMPTION.

| Models | Difficulty level of the tasks | Weighted score | Frequency (%) | Mean memory(kb) | Mean time(ms) |
|---|---|---|---|---|---|
| ChatGPT | Easy | 8.86 | 100 | 56 | 115 |
| | Medium | 11.64 | 54.5 | 3250 | 210 |
| | Hard | 2.95 | 11.1 | 100 | 124 |
| Deepseek | Easy | 7.87 | 88.8 | 82 | 126 |
| | Medium | 3.72 | 18.1 | 1650 | 545 |
| | Hard | 0 | 0 | 0 | 0 |

### B. Memory and time analysis

Fig. 4 presents a comparative analysis of execution time between GPT and DeepSeek for each programming task and individual problems. DeepSeek experiences a significant outlier at question 21, where the response time exceeds 4000 ms, which is substantially higher than any other recorded value. Overall, ChatGPT has shown to have shorter execution time in comparison to DeepSeek in solving coding tasks.

ChatGPT demonstrated more optimized memory usage and faster execution times in easy and medium level tasks as shown in Fig 5. While DeepSeek technically correct in some cases, it exhibited higher memory consumption, suggesting potential inefficiencies in code optimization. Even Though ChatGPT often produced incorrect answers it managed to utilize less memory compared to DeepSeek, which exceeded memory limits or ran into compilation errors.

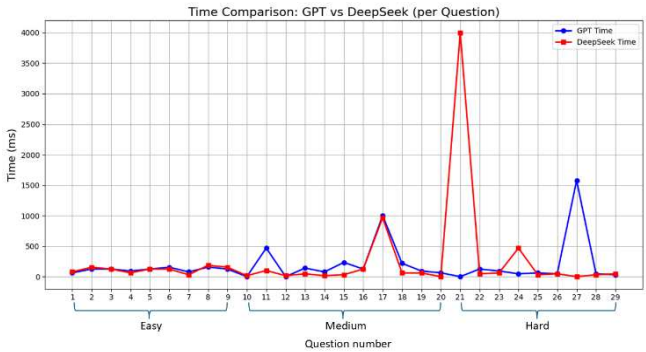

Fig 4. Time comparison for 29 programming tasks using ChatGPT o3-mini and DeepSeek-R1.

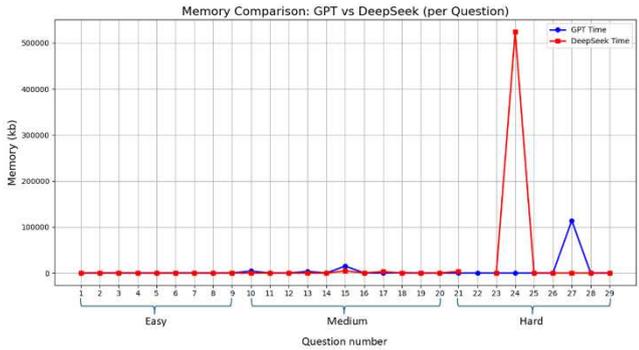

Fig 5. Memory usage comparison for 29 programming tasks using ChatGPT o3-mini and DeepSeek-R1.

## V. Discussion and Conclusion

This study compares ChatGPT o3-mini and DeepSeek-R1 in competitive programming. Our findings show that ChatGPT outperformed DeepSeek in terms of accepted solutions (i.e., correctness ratio).

For *easy* programming tasks, both models performed generally well, with high accuracy rates. This result suggests that LLMs may be relied on as proficient in handling straightforward programming problems, likely because such problems align closely with the training data they have been exposed to [7, 21].

Comparative analysis suggests that ChatGPT demonstrates superior performance to the R1 model in medium level programming tasks. This disparity may be attributed to the o3-mini model architecture, of which ChatGPT is a part, to benefit from more extensive pre-training on competitive programming data.

On the difficulty level of *hard* programming tasks from Codeforces, both models show a high threshold of struggling (i.e., 11% success rate for the o3-mini and 0% for the R1). The struggles of both models on hard tasks may imply the need for continued human expertise in solving advanced programming problems, as also discussed in [2]. Another implication from this finding calls for further need to optimize the LLM architecture to tackle high-complexity programming tasks, a capability we foresee emerging in the near term.

ChatGPT and DeepSeek manage memory and time efficiently for easy programming tasks without exceeding memory and time. For medium level tasks, DeepSeek had a lower memory consumption than ChatGPT, this could be due to DeepSeek's ability to reduce computation load through reinforcement learning based training [8]. However, despite its low memory usage, DeepSeek often struggled with optimization, occasionally producing incorrect solutions for medium level tasks. For hard programming tasks, both models face significant computational challenges. DeepSeek frequently exceeded memory limits, and encountered compilation errors, while ChatGPT also struggled with providing correct solutions to the hard programming tasks. The overall memory usage for ChatGPT is quite high compared to DeepSeek, one of the reasons could be ChatGPT uses all its parameters, i.e. 1.8 trillion parameters which makes it computationally intensive [3].

We've identified potential areas for implications for end-users. It's likely that zero-shot and single-shot prompting won't be sufficient for efficient solutions to medium and hard programming tasks. Therefore, we recommend that novice programmers explore more advanced prompting techniques to optimize for correctness, memory usage, and execution time from LLMs. Another area is the provision of code explainability by LLMs. This capability, as reported by [10], may enable lexical and semantic services equivalent to those offered by human experts and as a result enabling more opportunities to consider LLMs as AI-assisted learning tools.

### A. Study Limitations

The study comes with some limitations worth mentioning. First, our study relies on a single-shot nature from the LLMs responses. As no follow-up prompts were employed, any inaccuracies in the generated outputs were not refined. This might introduce inefficiencies, as LLM-assisted programming requires human intervention to ensure correctness [2]. Another fact which should also be considered is that for our experiment we have used ChatGPT o3-mini but for DeepSeek its more programming-refined version DeepSeek-Coder was not used, the performance for DeepSeek-Coder could have demonstrated better results than the R1. This opens up possibilities for future research. Another limitation is that only 29 programming tasks were tested, increasing the number and diversity of problem sets could help in better generalizing the results. Additionally, this study focuses on a single programming language, which may limit its applicability across different coding environments. Expanding the scope to multiple languages, such as Python, could provide broader insights. Furthermore, the performance of LLMs is highly dependent on prompt formulation. While a consistent prompt structure was used, exploring different prompts could offer deeper insights into model capabilities and response variability. As LLM models continue to evolve, further comparisons between different models will drive innovation and refinement in programming problems solving efficiency and computational optimization.

## References

[1] Z. Chen, J. Wang, M. Xia, K. Shigyo, D. Liu, R. Zhang, and H. Qu, "StuGPTViz: A visual analytics approach to understand student-ChatGPT interactions," *arXiv preprint arXiv:2407.12423*, 2024. [Online]. Available: https://arxiv.org/abs/2407.12423.

[2] V. Chugani, "DeepSeek vs. OpenAI: Comparing the New AI Titans," *DataCamp*, Feb. 6, 2025. [Online]. Available: https://www.datacamp.com/blog/deepseek-vs-openai.

[3] C. E. Coello, M. N. Alimam, and R. Kouatly, "Effectiveness of ChatGPT in coding: A comparative analysis of popular large language

models," *Digital*, vol. 4, no. 1, pp. 114–125, 2024. doi: 10.3390/digital4010005.

[4] *Dirox*, "DeepSeek vs ChatGPT vs Gemini: Choosing the Right AI for Your Needs," Feb. 20, 2025. [Online]. Available: https://dirox.com/post/deepseek-vs-chatgpt-vs-gemini-ai-comparison.

[5] E. Evstafev, "Token-Hungry, Yet Precise: DeepSeek R1 highlights the need for multi-step reasoning over speed in MATH," *arXiv preprint arXiv:2501.18576*, 2025. doi: 10.48550/arxiv.2501.18576.

[6] T. Gao, J. Jin, Z. T. Ke, and G. Moryoussef, "A Comparison of DeepSeek and Other LLMs," *arXiv preprint arXiv:2502.03688*, 2025. doi: 10.48550/arxiv.2502.03688.

[7] E. Gibney, "China's cheap, open AI model DeepSeek thrills scientists," *Nature*, pp. 13–14, 2025. doi: 10.1038/d41586-025-00229-6.

[8] S. Kapoor, "Large language models must be taught to know what they don't know," *arXiv preprint arXiv:2406.08391v2*, 2023. [Online]. Available: https://arxiv.org/html/2406.08391v2.

[9] G. Kaur, "DeepSeek vs ChatGPT: Comparing Features in 2025," *Cointelegraph*, 2025. [Online]. Available: https://cointelegraph.com/learn/articles/deepseek-vs-chatgpt.

[10] M. Kazemitabaar et al., "How novices use LLM-based code generators to solve CS1 coding tasks in a self-paced learning environment," *arXiv preprint arXiv:2309.14049*, 2023. doi: 10.48550/arXiv.2309.14049.

[11] M. Mirzayanov et al., "Codeforces as an educational platform for learning programming in digitalization," *Olympiads in Informatics*, pp. 133–142, 2020. doi: 10.15388/ioi.2020.10.

[12] S. Nguyen, H. M. Babe, Y. Zi, A. Guha, C. J. Anderson, & M. Q. Feldman, "How beginning programmers and code LLMs (mis)read each other," *arXiv preprint arXiv:3613904.3642706*, 2024. doi: 10.1145/3613904.3642706.

[13] A. Noriega, "Chinese AI makes a strong showing: DeepSeek-R1 outperforms ChatGPT in performance and efficiency," *Driving ECO*, Jan. 23, 2025. [Online]. Available: https://www.drivingeco.com/en/ia-china-pisa-fuerte-deepseek-r1-supera-chatgpt-rendimiento-eficiencia.

[14] OpenAI, "Hello GPT-4o," *OpenAI*, 2024. [Online]. Available: https://openai.com/index/hello-gpt-4o.

[15] OpenAI, "OpenAI o3-mini," *OpenAI*, 2025. [Online]. Available: https://openai.com/index/openai-o3-mini.

[16] S. Peng, E. Kalliamvakou, P. Cihon, and M. Demirer, "The impact of AI on developer productivity: Evidence from GitHub Copilot," *arXiv preprint arXiv:2302.06590*, 2023. doi: 10.48550/arxiv.2302.06590.

[17] L. Perez, "ChatGPT o3-mini-high: A leap forward in AI reasoning," *Neuroflash*, Feb. 5, 2025. [Online]. Available: https://neuroflash.com/blog/chatgpt-o3-mini-high.

[18] R. A. Poldrack, T. Lu, and G. Beguš, "AI-assisted coding: Experiments with GPT-4," *arXiv preprint arXiv:2304.13187*, 2023. [Online]. Available: https://arxiv.org/abs/2304.13187.

[19] B. Qureshi, "Exploring the use of ChatGPT as a tool for learning and assessment in undergraduate computer science curriculum: Opportunities and challenges," *arXiv preprint arXiv:2304.11214*, 2023. doi: 10.48550/arxiv.2304.11214.

[20] R. Ramler, M. Moser, L. Fischer, M. Nissl, & R. Heinzl, "Industrial experience report on AI-assisted coding in professional software development," in *Proc. 1st Int. Workshop Large Language Models for Code (LLM4Code '24)*, 2024, pp. 1–7. doi: 10.1145/3643795.3648377.

[21] Z. Tian and J. Chen, "Test-case-driven programming understanding in large language models for better code generation," *arXiv preprint arXiv:2309.16120*, 2023. doi: 10.48550/arxiv.2309.16120.

[22] T. Y. Yeh, K. Tran, G. Gao, T. Yu, W. O. Fong, T. Y. & Chen, "Bridging novice programmers and LLMs with interactivity," in *Proc. 56th ACM Tech. Symp. Comput. Sci. Educ. (SIGCSETS 2025)*, 2025, pp. 1295–1301. doi: 10.1145/3641554.3701867.

[23] Q. Zhu, D. Guo, Z. Shao, D. Yang, P. Wang, R. Xu ... & W. Liang, "DeepSeek-Coder-V2: Breaking the barrier of closed-source models in code intelligence," *arXiv preprint arXiv:2406.11931*, [n.d.]. [Online]. Available: https://arxiv.org/pdf/2406.11931.

[24] S. Yadav, A. M. Qureshi, A. Kaushik, S. Sharma, R. Loughran, S. Kazhuparambil, et al., "From idea to implementation: Evaluating the influence of large language models in software development—An opinion paper," *arXiv preprint arXiv:2503.07450*, 2025.

[25] A. Ni, P. Yin, Y. Zhao, M. Riddell, T. Feng, R. Shen, et al., "L2ceval: Evaluating language-to-code generation capabilities of large language models," *Trans. Assoc. Comput. Linguist.*, vol. 12, pp. 1311–1329, 2024.